\documentclass[a4paper]{article} 
\usepackage{amsmath,amsfonts,amsthm,eucal}
\numberwithin{equation}{section}

\newtheorem{Th}{Theorem}[section]
\newtheorem{Lem}[Th]{Lemma}
\newtheorem{Prop}[Th]{Proposition}
\newtheorem{Cor}[Th]{Corollary}

\newtheorem{Def}[Th]{Definition}

\theoremstyle{remark}
\newtheorem*{Rem}{Remark}

\newcommand{\nc}{\newcommand}
\nc{\ANN}{annihilat}
\nc{\ASS}{associat}
\nc{\AUT}{automorphism}
\nc{\BOG}{Bogoliubov }
\nc{\BP}{basis projection}
\nc{\CMP}[1]{{\em Commun. Math. Phys.} {\bf#1}}
\nc{\COR}{correspond}
\nc{\CR}{commutation relation}
\nc{\DEC}{decompos}
\nc{\DIM}{dimension}
\nc{\END}{endomorphism}
\nc{\HS}{Hilbert--Schmidt}
\nc{\HSP}{Hilbert space}
\nc{\IFF}{if and only if }
\nc{\IMP}{implement}
\nc{\IRR}{irreducib}
\nc{\ISM}{isometr}
\nc{\ISO}{isomorph}
\nc{\ONB}{orthonormal basis}
\nc{\OP}{operator}
\nc{\QEQ}{quasi--equivalen}
\nc{\QF}{quadratic form}
\nc{\REP}{representation}
\nc{\RUI}{Ruijsenaars}
\nc{\SG}{semigroup}
\nc{\SYM}{symmetr}
\nc{\WRT}{with respect to }

\nc{\DMO}{\DeclareMathOperator}
\DMO{\IND}{ind}          
\DMO{\RAN}{ran}
\DMO{\RE}{Re}          
\DMO{\TR}{tr}          

\nc{\0}{\mspace{-1.5mu}\circ\mspace{-0.5mu}}
\nc{\1}{{\boldsymbol1}}                   
\nc{\2}{{\frac{1}{2}}}
\nc{\4}[1]{{\overline{#1}}}  
\nc{\+}{^\dagger}             
\nc{\ABS}[1]{\lvert{#1}\rvert}                  
\nc{\BB}{{\mathfrak B}}
\nc{\CC}{{\mathbb C}}            
\nc{\CK}{{\mathcal C}(\KK,\gamma)}    
\nc{\CKO}{{\mathcal C}(\KK^0,\gamma)}    
\nc{\DD}{{\mathcal D}} 
\nc{\DEF}{\equiv}
\nc{\EE}{{\mathcal E}}      
\nc{\FF}{{\mathcal F}}      
\nc{\FK}{{\FF_s(\KK_1)}} 
\nc{\HH}{{\mathcal H}}                    
\nc{\IF}{\Leftrightarrow}
\nc{\KK}{{\mathcal K}}                  
\nc{\NN}{{\mathbb N}}                  
\nc{\NORM}[1]{\lVert{#1}\rVert}                  
\nc{\PIP}{{\pi_{P_1}}}
\nc{\PP}{{\mathcal P}}                  
\nc{\R}[1]{{\varrho_{#1}}}   
\nc{\RR}{{\mathbb R}}      
\nc{\SP}[2]{{{\mathcal S}_{{#1}}^{#2}({\KK},\gamma)}}
\nc{\SSS}{{\mathfrak S}}
\nc{\WK}{{\WW(\KK,\gamma)}}      
\nc{\WO}[1]{{\::\!{#1}\!:\:}} 
\nc{\WW}{{\mathcal W}}

\begin{document}
\title{${\mathcal O}_\infty$ Realized on Bose Fock Space}
\author{Carsten Binnenhei\thanks{Supported by the Deutsche
    Forschungsgemeinschaft (Sfb 288 ``Differentialgeometrie und
    Quantenphysik'')}\ \thanks{E--mail: binnenhe@physik.fu-berlin.de}
  \\[1ex]Institut f\"ur Theoretische Physik der Freien Universit\"at,\\
  Arnimallee 14, 14195 Berlin, Germany}
\date{December 1996}
\maketitle
\begin{abstract}
  We study the \SG\ of \BOG\END s of the canonical \CR s which give
  rise to \REP s of the Cuntz algebra ${\mathcal O}_\infty$ on Fock
  space and describe the \COR ing Cuntz algebra generators in detail.
\end{abstract}
\section{Introduction}
\label{sec:INTRO}
The appearance of the Cuntz algebras ${\mathcal O}_d$ \cite{C} is a
generic feature of quantum field theory. This fact has been
discovered, within the algebraic approach \cite{H}, by Doplicher and
Roberts~\cite{DR} who \ASS ed with each localized morphism $\R{}$ of
\DIM\ $d$ and obeying permutation group statistics a multiplet
$\Psi_1,\dots,\Psi_d$ of local field \OP s, acting on a \HSP\ which
contains each superselection sector with some multiplicity, such that
Cuntz' relations hold
\begin{equation}
  \label{CUNTZ0}
  \Psi_j^*\Psi_k=\delta_{jk}\1,\quad \sum\limits_j\Psi_j\Psi_j^*=\1  
\end{equation}
and such that, for any local observable $A$, 
\begin{equation}
  \label{IMP0}
  \R{}(A)=\sum_j\Psi_jA\Psi_j^*.
\end{equation}
Regarding the $\Psi_j$ as an \ONB\ for the closure $H(\R{})$ of their
linear span, one says that `$\R{}$ is \IMP ed by the \HSP\ $H(\R{})$
of \ISM ies'.

These observations motivated a study of \END s of the CAR algebra
which can be \IMP ed by \HSP s of \ISM ies on Fock space \cite{CB}.
Here we present a similar analysis for \END s of the canonical
commutation relations (CCR) or, more precisely, of the Weyl relations.
As in \cite{CB}, we find it convenient to use Araki's `selfdual'
formulation \cite{A71,A82} which is briefly introduced in
Sect.~\ref{sec:BAS}. We discuss a class of natural \END s of the CCR
algebra (`\BOG\END s') in Sect.~\ref{sec:IMP}. As a generalization of
Shale's condition for \AUT s \cite{S}, we state a necessary and
sufficient condition for \BOG\END s to be \IMP able in a fixed Fock
\REP. The derivation of this result is based on known criteria for
\QEQ ce of quasi--free states \cite{A71,vD,A82} and can, at one point,
be reduced to the CAR case, by using an inequality due to Araki and
Yamagami \cite{AY}.  

The topological \SG\ of \IMP able \END s is the subject of
Sect.~\ref{sec:SG}. It can be written as a product of a subgroup
consisting of \AUT s which are close to the identity, and the
sub--\SG\ of \END s which leave the given Fock state invariant. This
\DEC ition enables us to determine the connected components of the
\SG. It also plays a role in Sect.~\ref{sec:CON} which is
concerned with the construction of orthonormal bases for the \HSP s
$H(\R{})$. These \HSP s themselves carry the structure of \SYM ic
Fock spaces and thus are, for genuine \END s, infinite--\DIM al. The
C*--algebra generated by a single $H(\R{})$ is ${\mathcal O}_\infty$.
Implementers are constructed by an adaptation of Ruijsenaars' formulas
for unitary \IMP ers of \AUT s \cite{R78}. They can be written as
products of certain \ISM ies belonging to the commutant of the range
of $\R{}$ times a Wick ordered exponential of an expression which is
bilinear in creation and \ANN ion \OP s. The connection with the
aforementioned product \DEC ition is that, roughly speaking, the first
factor carries the exponential term, whereas the second is responsible
for the additional \ISM ies. The proof of completeness of \IMP ers
can thereby be reduced to the case of \END s which leave the given
Fock state invariant.
\section{Basic Notions}
\label{sec:BAS}
Let $\KK^0$ be an infinite--\DIM al complex linear space, equipped
with a nondegenerate hermitian sesquilinear form $\gamma$ and an
antilinear involution $f\mapsto f^*$, such that
$$\gamma(f^*,g^*)=-\gamma(g,f),\qquad f,g\in\KK^0.$$ (The reader who
is unfamiliar with Araki's approach~\cite{A71,A82} should think of
$\KK^0$ as being the complexification of the real linear space
$\RE\KK^0\DEF\{f\in\KK^0\ |\ f^*=f\}$, together with its canonical
conjugation.
$-i\gamma$ should be viewed as the sesquilinear extension of a
nondegenerate symplectic form on $\RE\KK^0$. Hopefully, the reader
will not be confused in the following by the appearance of too many
stars with different meanings.) The {\em selfdual CCR algebra}
$\CKO$~\cite{A71,A82} over $(\KK^0,\gamma)$ is the simple *--algebra
which is generated by $\1$ and elements $f\in\KK^0$, subject to the
\CR
\begin{equation}
  \label{CCR}
  f^*g-gf^*=\gamma(f,g)\1,\quad f,g\in\KK^0.
\end{equation}
We henceforth assume the existence of a distinguished Fock state
$\omega_{P_1}$.  
Here $P_1$ is a {\em\BP} of $\KK^0$, i.e.\ a linear \OP, defined on
the whole of $\KK^0$, which satisfies
\begin{equation}
  \label{BP}
  \begin{aligned}
    P_1^2&=P_1, &\qquad \gamma(f,P_1g)&=\gamma(P_1f,g),\\ 
    P_1f+P_1(f^*)^*&=f, &\qquad\gamma(f,P_1f)&>0\text{ if }P_1f\not=0
  \end{aligned}
\end{equation}
for $f,g\in\KK^0$. Let
$$P_2\DEF\1-P_1,\quad C\DEF P_1-P_2,\quad\langle{f,g}\rangle
\DEF\gamma(f,Cg).$$ The positive definite inner product $\langle{\ ,\ 
  }\rangle$ turns $\KK^0$ into a pre--\HSP. We assume its completion
$\KK$ to be separable. By continuity, the involution $*$ extends to a
conjugation on $\KK$, $P_1$ and $P_2$ to orthogonal projections, $C$
to a self--adjoint unitary, and $\gamma$ to a nondegenerate hermitian
form. These extensions will be denoted by the same symbols. Setting
$$\KK_n\DEF P_n(\KK),\qquad n=1,2,$$ we get a direct sum \DEC ition
$\KK=\KK_1\oplus\KK_2$ which is orthogonal \WRT both $\gamma$ and
$\langle\ ,\ \rangle$. The following notations will frequently be used
for $A\in\BB(\KK)$, where $\BB(\KK)$ is the algebra of bounded linear
\OP s on $\KK$:
\begin{align*}
  A_{mn} & \DEF P_mAP_n,\quad m,n=1,2,\\
  A\+    & \DEF CA^*C,\\
  \4{A}f & \DEF A(f^*)^*,\quad f\in\KK.
\end{align*}
The components $A_{mn}$ of $A$ are regarded as \OP s from $\KK_n$ to
$\KK_m$, and $A$ will sometimes be written as a matrix
$\bigl(\begin{smallmatrix}A_{11}&A_{12}\\ 
  A_{21}&A_{22}\end{smallmatrix}\bigr)$. $A\+$ is the adjoint of $A$
relative to $\gamma$, whereas $A^*$ is the \HSP\ adjoint. $\4{A}$ may
be viewed as the complex conjugate of $A$. Thus one has relations like
$$\4{P_2}=P_1=P_1\+=P_1^*,\ \4{C}=-C,\ {A_{12}}\+={A\+}_{21}=
-{A_{12}}^*,\ \4{A_{11}}=\4{A}_{22}\text{\quad etc.}$$ 
The {\em Fock state} $\omega_{P_1}$ is the unique state\footnote{A
  state $\omega$ over $\CK$ is a linear functional with $\omega(\1)=1$
  and $\omega(A^*A)\geq0,\ A\in\CK$.} which is \ANN ed by all
$f\in\RAN P_2$:
$$\omega_{P_1}(f^*f)=0\text{ if }P_1f=0.$$ (In the conventional
setting mentioned above, $\omega_{P_1}$ is the Fock state \COR ing to
the complex structure $iC$ on $\RE\KK$.) Let $\FK$ be the \SYM ic Fock
space over $\KK_1$ and let $\DD$ be the dense subspace of algebraic
tensors. A GNS \REP\ $\PIP$ for $\omega_{P_1}$ is provided by
$$\PIP(f)=a^*(P_1f)+a\bigl(P_1(f^*)\bigr),\quad f\in\KK$$ where
$a^*(g)$ and $a(g)$, $g\in\KK_1$, are the usual creation and \ANN ion
\OP s on $\DD$. The cyclic vector inducing the state $\omega_{P_1}$ is
$\Omega_{P_1}$, the Fock vacuum. The \OP s $\PIP(a)$, $a\in\CK$, have
invariant domain $\DD$, are closable, and $\PIP(a^*)\subset\PIP(a)^*$.
In particular, if $f\in\RE\KK$, then $\PIP(f)$ is essentially
self--adjoint on $\DD$, and the unitary {\em Weyl \OP} $w(f)$ is
defined as the exponential of the closure of $i\PIP(f)$. Its vacuum
expectation value is
$$\omega_{P_1}(w(f))\DEF\langle{\Omega_{P_1},
  w(f)\Omega_{P_1}}\rangle=e^{-\2\NORM{P_1f}^2},$$
and the {\em Weyl relations} hold
$$w(f)w(g)=e^{-\2\gamma(f,g)}w(f+g),\quad f,g\in\RE\KK.$$
The Weyl \OP s generate a simple C*--algebra $\WK$ which acts
\IRR ly on $\FK$. If $\HH$ is a subspace of $\KK$ with $\HH=\HH^*$,
then the C*--algebra generated by all $w(f)$ with $f\in\RE\HH$ is
denoted by $\WW(\HH)$. If $\HH_0$ is the orthogonal complement of
$\HH$ \WRT $\gamma$, then {\em duality} holds \cite{A63,A71,A82}:
\begin{equation}
  \label{DUAL}
  \WW(\HH)'=\WW(\HH_0)''
\end{equation}
(a prime denotes the commutant). 
\begin{Lem}
  \label{lem:AFF}
  For $f\in\KK$, let $\HH_f$ be the subspace spanned by $f$ and $f^*$.
  Then the closure of $\PIP(f)$ is affiliated with $\WW(\HH_f)''$.
\end{Lem}
\begin{proof}
  Let $T$ be the closure of $\PIP(f)$, with domain $D(T)$. We have to
  show that, for any $A\in\WW(\HH_f)'$
  $$A(D(T))\subset D(T),\qquad AT=TA\text{ on }D(T).$$ Now by virtue
  of the CCR \eqref{CCR}, $\NORM{T\phi}^2=\NORM{T^*\phi}^2
  +\gamma(f,f)\NORM{\phi}^2$ for $\phi\in\DD$.  Hence, for a given
  Cauchy sequence $\phi_n\in\DD$, $T\phi_n$ converges \IFF $T^*\phi_n$
  does. This implies that
  $$D(T)=D(T^*).$$ Let $f^\pm\in\RE\HH_f$ be defined as
  $f^+\DEF\2(f+f^*),\ f^-\DEF\frac{i}{2}(f-f^*)$, and let $T^\pm$ be
  the (self--adjoint) closure of $\PIP(f^\pm)$. We claim that
  $$D(T)=D(T^+)\cap D(T^-),\qquad T=T^+-iT^-\text{ on }D(T).$$ For if
  $\phi\in D(T)$, then there exists a sequence $\phi_n\in\DD$
  converging to $\phi$ such that $\PIP(f)\phi_n$ and $\PIP(f^*)\phi_n$
  converge. Thus $\phi$ belongs to the domain of the closure of
  $\PIP(f^\pm)$. Conversely, if $\phi\in D(T^+)\cap D(T^-)$, then
  there exists a sequence $\phi_n\in\DD$ converging to $\phi$ such
  that both $\PIP(f+f^*)\phi_n$ and $\PIP(f-f^*)\phi_n$ converge
  (cf.~\cite{R78}).  Therefore $\PIP(f)\phi_n$ is also convergent,
  i.e.\ $\phi$ is contained in $D(T)$, and $T\phi=(T^+-iT^-)\phi$.
  
  Now if $A\in\WW(\HH_f)'$, then $A$ commutes with the one--parameter
  unitary groups $w(tf^\pm)=\exp(itT^\pm)$. As a consequence, $A$
  leaves $D(T^\pm)$ invariant and commutes with $T^\pm$ on $D(T^\pm)$.
  It follows that $A(D(T))\subset D(T)$ and $AT=TA$ on $D(T)$ as was
  to be shown.
\end{proof}
\section{Implementability of Endomorphisms}
\label{sec:IMP}
{\em\BOG\END s} are the unital *--\END s of $\CK$ which map $\KK$,
viewed as a subspace of $\CK$, into itself. They are completely
determined by their restrictions to $\KK$ which are called {\em\BOG\OP
  s}. Hence $V\in\BB(\KK)$ is a \BOG\OP\ \IFF it commutes with complex
conjugation and preserves the hermitian form $\gamma$\footnote{We may
  disregard unbounded \BOG\OP s $V$ (defined on $\KK^0$) since the
  topologies induced by the \COR ing states $\omega_{P_1}\0\R{V}$ on
  $\KK^0$ differ from the one induced by $\omega_{P_1}$. Hence these
  states cannot be \QEQ t to $\omega_{P_1}$ (cf.~\cite{A71,A82}),
  and $\R{V}$ cannot be \IMP ed.}.
\BOG\OP s form a unital \SG\ denoted by
$$\SP{}{}\DEF\{V\in\BB(\KK)\ |\ \4{V}=V,\ V\+V=\1\}.$$ Each
$V\in\SP{}{}$ extends to a unique \BOG\END\ of $\CK$ and to a unique
*--\END\ of $\WK$. By abuse of notation, both \END s are denoted by
$\R{V}$, so that $\R{V}(f)=Vf,\ f\in\KK$, and $\R{V}(w(g))=w(Vg),\ 
g\in\RE\KK$.

The condition $V\+V=\1$ entails that $V$ is injective and $V^*$
surjective; hence $\RAN V$ is closed, and $V$ is a semi--Fredholm
\OP~\cite{K}. We claim that the Fredholm index $-\IND V=\dim\ker V\+$
cannot be odd, in contrast to the CAR case~\cite{CB}. For let
$f\in\ker V\+$ such that $0=\gamma(f,g)\equiv\langle{f,Cg}\rangle\ 
\forall g\in\ker V\+$. Then $f\in(C\ker V\+)^\perp=(\ker
V^*)^\perp=\RAN V$, but $\RAN V\cap\ker V\+=\{0\}$ due to $V\+V=\1$,
so $f$ has to vanish. This shows that the restriction of $\gamma$ to
$\ker V\+$ stays nondegenerate. It follows that $\dim\ker V\+$ cannot
be odd (there is no nondegenerate symplectic form on an odd--\DIM al
space).

On the other hand, each even number (and $\infty$) occurs as $\dim\ker
V\+$ for some $V$. Hence we have an epimorphism of \SG s
$$\SP{}{}\to\NN\cup\{\infty\},\quad V\mapsto-\2\IND
V=\2\dim\ker V\+$$
($0\in\NN$ by convention). Let
$$\SP{}{n}\DEF\{V\in\SP{}{}\ |\ \IND V=-2n\},\quad
n\in\NN\cup\{\infty\}.$$ 
$\SP{}{0}$ is the group of \BOG\AUT s (\ISO ic to the symplectic group
of $\RE\KK$). It acts on $\SP{}{}$ by left multiplication. Analogous
to the CAR case, the orbits under this action are the subsets
$\SP{}{n}$, and the stabilizer of $V\in\SP{}{n}$ is \ISO ic to the
symplectic group Sp($n$).

We are interested in \END s $\R{V}$ which can be \IMP ed by \HSP s of
\ISM ies on $\FK$. This means that there exist \ISM ies $\Psi_j$ on
$\FK$ which fulfill the Cuntz algebra relations \eqref{CUNTZ0} and
\IMP\ $\R{V}$ according to \eqref{IMP0}
$$\R{V}(w(f))=\sum_j\Psi_jw(f)\Psi_j^*,\quad f\in\RE\KK.$$ As
explained in~\cite{CB}, such \ISM ies exist \IFF $\R{V}$, viewed as a
\REP\ of $\WK$ on $\FK$, is \QEQ t to the defining (Fock) \REP.

To study $\R{V}$ as a \REP, for fixed $V\in\SP{}{}$, let us \DEC e it
into cyclic sub\REP s.  Let $e_1,e_2,\dotsc$ be an \ONB\ in
$\KK_1\cap\ker V\+$ and let $\alpha=(\alpha_1,\dots,\alpha_l)$ be a
multi--index with $\alpha_j\leq\alpha_{j+1}$. Such $\alpha$ has the
form 
\begin{equation}
  \label{ALPHA}
  \alpha=(\underbrace{\alpha'_1,\dots,\alpha'_1}_{l_1},
  \underbrace{\alpha'_2,\dots,\alpha'_2}_{l_2},\dots,
  \underbrace{\alpha'_r,\dots,\alpha'_r}_{l_r})
\end{equation}
with $\alpha'_1<\alpha'_2<\dotsb<\alpha'_r$ and $l_1+\dots+l_r=l$. Let
\begin{align*}
  \phi_\alpha&\DEF(l_1!\dotsm l_r!)^{-\2}a^*(e_{\alpha_1})\dotsm
    a^*(e_{\alpha_l})\Omega_{P_1},\\ 
  \FF_\alpha &\DEF \4{\WW(\RAN V)\phi_\alpha},\\ 
  \pi_\alpha &\DEF \R{V}|_{\FF_\alpha}.
\end{align*}
\begin{Lem}\label{lem:CYC}
  $\R{V}=\oplus_\alpha\pi_\alpha$, where the sum extends over all
  multi--indices $\alpha$ as above, including $\alpha=0$
  ($\phi_0\DEF\Omega_{P_1}$). Each
  $(\pi_\alpha,\FF_\alpha,\phi_\alpha)$ is a GNS \REP\ for
  $\omega_{P_1}\0\R{V}$ (regarded as a state over $\WK$).
\end{Lem}
\begin{proof}
  By definition, the $\phi_\alpha$ constitute an \ONB\ for
  $\FF_s(\KK_1\cap\ker V\+)$, and
  $(\pi_\alpha,\FF_\alpha,\phi_\alpha)$ is a cyclic \REP\ of $\WK$.
  Since the closures of $a^*(e_j)$ and $a(e_j)$ are affiliated with
  $\WW(\ker V\+)''=\WW(\RAN V)'$ (see Lemma~\ref{lem:AFF} and
  \eqref{DUAL}), there holds for $f\in\RE\KK$, with
  $N_\alpha\DEF(l_1!\dotsm l_r!)^{-1}$
  \begin{equation*}
    \begin{split}
      \langle{\phi_\alpha,\pi_\alpha(w(f))\phi_\alpha}\rangle &=
      N_\alpha\langle{a^*(e_{\alpha_1})\dotsm
        a^*(e_{\alpha_l})\Omega_{P_1}, w(Vf)a^*(e_{\alpha_1})\dotsm
        a^*(e_{\alpha_l})\Omega_{P_1}}\rangle\\ 
      &=N_\alpha\langle\Omega_{P_1},
      w(Vf)\underbrace{a(e_{\alpha_l})\dotsm
        a(e_{\alpha_1})a^*(e_{\alpha_1})\dotsm a^*(e_{\alpha_l})
        \Omega_{P_1}}_{N_\alpha^{-1}\Omega_{P_1}}\rangle\\ &=
      \langle{\Omega_{P_1},w(Vf)\Omega_{P_1}}\rangle.
    \end{split}
  \end{equation*}
  This proves that $(\pi_\alpha,\FF_\alpha,\phi_\alpha)$ is a GNS
  \REP\ for $\omega_{P_1}\0\R{V}$. Similarly, one finds that
  $\langle{\phi_\alpha,w(Vf)\phi_{\alpha'}}\rangle=0$ for
  $\alpha\neq\alpha'$, so the $\FF_\alpha$ are mutually orthogonal.

  It remains to show that $\oplus_\alpha\FF_\alpha=\FK$. We claim that
  $\FF_0$ equals $\FF_s(\4{\RAN P_1V})$, the \SYM ic Fock space over
  the closure of $\RAN P_1V$. The inclusion $\FF_0\subset\FF_s(\4{\RAN
    P_1V})$ holds because vectors of the form $w(Vf)\Omega_{P_1}=\exp
  i\bigl(a^*(P_1Vf)+a(P_1Vf)\bigr)\Omega_{P_1}\in\FF_s(\4{\RAN P_1V})$
  are total in $\FF_0$. The converse inclusion may be proved
  inductively. Assume that $a^*(g_1)\dotsm a^*(g_m)\Omega_{P_1}$ is
  contained in $\FF_0$ for all $m\leq n,\ g_1,\dots,g_m\in\RAN P_1V$.
  Then, for $f\in V(\RE\KK)$ and $g_1,\dots,g_n\in\RAN P_1V$,
  $\tfrac{1}{i}\tfrac{w(tf)-\1}{t}a^*(g_1)\dotsm a^*(g_n)\Omega_{P_1}$
  has a limit $a^*(P_1f)a^*(g_1)\dotsm a^*(g_n)\Omega_{P_1}+
  a(P_1f)a^*(g_1)\dotsm a^*(g_n)\Omega_{P_1}$ in $\FF_0$ as
  $t\searrow0$. By assumption, the second term lies in $\FF_0$, and so
  does the first. Since each $g\in\RAN P_1V$ is a linear combination
  of such $P_1f$, it follows that $a^*(g_1)\dotsm
  a^*(g_{n+1})\Omega_{P_1}$ is contained in $\FF_0$ for arbitrary
  $g_j\in\RAN P_1V$, and, by induction, for arbitrary $n\in\NN$. But
  such vectors span a dense subspace in $\FF_s(\4{\RAN P_1V})$, so
  $\FF_0=\FF_s(\4{\RAN P_1V})$ as claimed.

  Finally, $\KK_1\cap\ker V\+$ equals $\ker V^*P_1$, where $V^*P_1$ is
  regarded as an \OP\ from $\KK_1$ to $\KK$. Thus we have
  $\KK_1=\4{\RAN P_1V}\oplus(\KK_1\cap\ker V\+)$ and
  $\FK\cong\FF_0\otimes\FF_s(\KK_1\cap\ker V\+)$. Under this \ISO ism,
  $\FF_\alpha$ is identified with $\FF_0\otimes(\CC\phi_\alpha)$.
  Since the $\phi_\alpha$ form an \ONB\ for $\FF_s(\KK_1\cap\ker
  V\+)$, the desired result $\oplus_\alpha\FF_\alpha=\FK$ follows.
\end{proof}
As a consequence, the \REP\ $\R{V}$ is \QEQ t to the GNS \REP\ 
associated with the quasi--free state $\omega_{P_1}\0\R{V}$. So
$\R{V}$ is \IMP able \IFF $\omega_{P_1}\0\R{V}$ and $\omega_{P_1}$
(i.e.\ their GNS \REP s) are \QEQ t. Now the two--point function of
$\omega_{P_1}\0\R{V}$ (as a state over $\CK$) is given by
$$\omega_{P_1}\0\R{V}(f^*g)=\gamma(f,Sg)=\langle{f,\tilde{S}g}\rangle,
\quad f,g\in\KK,$$ with
$$S\DEF V\+P_1V,\qquad\tilde{S}\DEF V^*P_1V.$$ The latter \OP s
contain valuable information about $\omega_{P_1}\0\R{V}$. For example,
it can be shown (cf.~\cite{MV}) that $\omega_{P_1}\0\R{V}$ is a {\em
  pure} state over $\WK$ \IFF $S$ is a \BP, that is, \IFF $S$ is
idempotent (the remaining conditions in \eqref{BP} are automatically
fulfilled). This is further equivalent to $[P_1,VV\+]=0$, by the
following chain of equivalences:
$$\begin{array}{rcll}
  S^2=S &\IF& 0=S\4{S} & (\text{since }\4{S}=\1-S)\\
  &\IF& 0=V^*P_1VCV^*P_2V & \\
  &\IF& 0=P_1VCV^*P_2 & (\text{since }\RAN V^*P_2V=\RAN V^*P_2\\
  &&& \text{ and }\ker V^*P_1V=\ker P_1V)\\
  &\IF& 0=P_1VV\+P_2 & \\
  &\IF& 0=[P_1,VV\+].&
\end{array}$$
On the other hand, the criterion for \QEQ ce of quasi--free states, in
the form given by Araki and Yamagami~\cite{A82}, states that
$\omega_{P_1}\0\R{V}$ is \QEQ t to $\omega_{P_1}$ \IFF
$P_1-\tilde{S}^{\2}$ is a \HS\ \OP\ on $\KK$.  This condition can be
simplified in the present context, as the following result shows.
\begin{Th}\label{th:IMP}
  Let a \BOG\ \OP\ $V\in\SP{}{}$ be given. Then there exists a \HSP\ 
  of \ISM ies $H(\R{V})$ which \IMP s the \END\ $\R{V}$ in the Fock
  \REP\ determined by the \BP\ $P_1$ \IFF $[P_1,V]$ (or, equivalently,
  $V_{12}$) is a \HS\ \OP. The \DIM\ of $H(\R{V})$ is $1$ if\/ $\IND
  V=0$, otherwise $\infty$.
\end{Th}
\begin{proof}
  First note that $[P_1,V]=V_{12}-V_{21}=V_{12}-\4{V_{12}}$ is \HS\
  (HS) \IFF $V_{12}$ is. 

  By the preceding discussion, $\R{V}$ is \IMP able \IFF
  $P_1-\tilde{S}^{\2}$ is HS\@. In this case,
  $P_2(P_1-\tilde{S}^{\2})^2P_2=P_2\tilde{S}P_2= {V_{12}}^*V_{12}$ is
  of trace class, hence $V_{12}$ is HS\@.

  Conversely, assume $V_{12}$ to be HS\@. Let $V=V'\ABS{V}$ be the
  polar \DEC ition of $V$. Then $\ABS{V}=\4{\ABS{V}}$ is a bounded
  bijection with a bounded inverse, and $\ABS{V}-\1=
  (\ABS{V}^2-\1)(\ABS{V}+\1)^{-1}=(V^*-V\+)V(\ABS{V}+\1)^{-1}
  =2({V_{12}}^*+{V_{21}}^*)V(\ABS{V}+\1)^{-1}$ is HS\@. Thus, by a
  corollary~\cite{A82} of an inequality of Araki and
  Yamagami~\cite{AY}, $(\ABS{V}A\ABS{V})^\2-A^\2$ is HS for any
  positive $A\in\BB(\KK)$. Applying this to $A={V'}^*P_1V'$, we get
  that
  \begin{equation}
    \label{HS}
    \tilde{S}^\2-({V'}^*P_1V')^\2\text{ is HS\@.}
  \end{equation}
  Now $V'$ is an \ISM y with $\4{V'}=V'$, i.e.\ a CAR \BOG
  \OP~\cite{CB}. Since $[P_1,V]$ and $[P_1,\ABS{V}^{-1}]=
  \ABS{V}^{-1}\bigl[\ABS{V},P_1\bigr]\ABS{V}^{-1}=
  \ABS{V}^{-1}\bigl[\ABS{V}-\1,P_1\bigr]\ABS{V}^{-1}$ are HS, the same
  holds true for $[P_1,V']=[P_1,V\ABS{V}^{-1}]$. So $V'$ fulfills the
  \IMP ability condition for CAR \BOG \OP s derived in~\cite{CB}, and,
  as shown there, this forces $P_1-({V'}^*P_1V')^\2$ to be HS\@. This,
  together with \eqref{HS}, implies that $P_1-\tilde{S}^{\2}$ is HS as
  claimed.

  It remains to prove the statement about $\dim H(\R{V})$. Let
  $\tilde\varrho_V$ be the normal extension of $\R{V}$ to $\BB(\FK)$.
  Then $\BB(H(\R{V}))\cong\tilde\varrho_V(\BB(\FK))'=\R{V}(\WK)'
  =\WW(\RAN V)'=\WW(\ker V\+)''$. The latter (and hence $H(\R{V})$) is
  one--\DIM al if $\ker V\+=\{0\}$ and infinite--\DIM al if $\ker
  V\+\not=\{0\}$.
\end{proof}
\begin{Rem}
  Shale's original result~\cite{S} asserts that a \BOG\AUT\ $\R{V}$,
  $V\in\SP{}{0}$, is \IMP able \IFF $\ABS{V}-\1$ is HS\@.  This
  condition is equivalent to $[P_1,V]$ being HS, not only for
  $V\in\SP{}{0}$, but for all $V\in\SP{}{}$ with $-\IND V<\infty$.
  However, the two conditions are {\em not\/} equivalent for
  $V\in\SP{}{\infty}$, as the following example shows. Let
  $\KK_1=\HH\oplus\HH'$ be a \DEC ition into infinite--\DIM al
  subspaces. Choose an \OP\ $V_{12}$ from $\KK_2$ to $\HH$ with
  $\TR\ABS{V_{12}}^4<\infty$, but $\TR\ABS{V_{12}}^2=\infty$. Let
  $V_{21}\DEF\4{V_{12}}$ and $\ABS{V_{11}}\DEF(P_1+
  \ABS{V_{21}}^2)^\2$.  Choose an \ISM y $v_{11}$ from $\KK_1$ to
  $\HH'$ and set $V_{11}\DEF v_{11}\ABS{V_{11}},\ 
  V_{22}\DEF\4{V_{11}}$. These components define a \BOG \OP\ 
  $V\in\SP{}{\infty}$ (cf.\ \eqref{REL1}--\eqref{REL4} below) which
  violates the condition of Theorem~\ref{th:IMP}. But it fulfills
  Shale's condition since $\ABS{V}^2-\1=2(\ABS{V_{12}}^2+
  \ABS{V_{21}}^2)$ is HS and since $\ABS{V}-\1=(\ABS{V}^2-\1)
  (\ABS{V}+\1)^{-1}$.
\end{Rem}
Let $V\in\SP{}{}$ with $V_{12}$ compact. Due to stability under
compact perturbations \cite{K}, $V_{11}$ and $V_{22}=\4{V_{11}}$ are
semi--Fredholm with
\begin{equation}
  \label{IND}
  \IND V_{11}=\IND V_{22}=\2\IND V.
\end{equation}
We will occasionally use the relation $V\+V=\1$ componentwise:
\begin{subequations}\label{REL}
  \begin{align}
    {V_{11}}^*V_{11}-{V_{21}}^*V_{21} &= P_1, \label{REL1} \\
    {V_{22}}^*V_{22}-{V_{12}}^*V_{12} &= P_2, \label{REL2} \\
    {V_{11}}^*V_{12}-{V_{21}}^*V_{22} &= 0,   \label{REL3} \\
    {V_{22}}^*V_{21}-{V_{12}}^*V_{11} &= 0.   \label{REL4}
  \end{align}
\end{subequations}
Since $V_{11}$ is injective by \eqref{REL1} and has closed range, we
may define a bounded \OP\ ${V_{11}}^{-1}$ as the inverse of ${V_{11}}$
on $\RAN{V_{11}}$ and as zero on $\ker {V_{11}}^*$ (the same applies
to $V_{22}$). These \OP s will be needed later. Note that $\dim\ker
{V_{11}}^*=\2\IND V$.
\section{On the Semigroup of Implementable Endomorphisms}
\label{sec:SG}
According to Theorem~\ref{th:IMP}, the \SG\ of \IMP able \BOG\END s is
\ISO ic to the following \SG\ of \BOG\OP s:
$$\SP{P_1}{}\DEF\{V\in\SP{}{}\ |\ V_{12}\text{ is \HS}\}.$$
$\SP{P_1}{}$ is a topological \SG\ \WRT the metric $d_{P_1}(V,V')
\DEF\NORM{V-V'}+\NORM{V_{12}-V'_{12}}_{\text{HS}}$, where $\NORM{\ 
  }_{\text{HS}}$ denotes \HS\ norm. It contains the closed sub--\SG\ 
of diagonal \BOG\OP s
$$\SP{\text{diag}}{}=\{V\in\SP{}{}\ |\ [P_1,V]=0\}$$ 
which is \ISO ic to the \SG\ of \ISM ies of the \HSP\ $\KK_1$, via the
map $V\mapsto V_{11}$. The Fredholm index yields a \DEC ition 
$$\SP{P_1}{}=\bigcup_{n\in\NN\cup\{\infty\}}\SP{P_1}{n},\qquad
\SP{P_1}{n}\DEF\SP{P_1}{}\cap\SP{}{n}.$$ The group $\SP{P_1}{0}$ is
usually called the {\em restricted symplectic group} \cite{S,Se}. It
has a natural normal subgroup
$$\SP{\text{HS}}{}\DEF\{V\in\SP{}{}\ |\ V-\1\text{ is \HS}\}
\subset\SP{P_1}{0}.$$ We will eventually show that each
$V\in\SP{P_1}{}$ can be written as a product $V=UW$ with
$U\in\SP{\text{HS}}{}$ and $W\in\SP{\text{diag}}{}$. Assume that such
$U$ and $W$ exist. Then $P_V\DEF UP_1U\+$ is a \BP\ such that
\begin{equation}
  \label{PV}
  P_1-P_V\text{ is \HS},\qquad V\+P_VV=P_1,
\end{equation}
so the \COR ing Fock state $\omega_{P_V}$ is unitarily equivalent to
$\omega_{P_1}$ and fulfills $\omega_{P_V}\0\R{V}=\omega_{P_1}$. In
order to construct such \BP s, let us investigate the set $\PP_{P_1}$
of \BP s of $\KK$ which differ from $P_1$ only by a \HS\ \OP. Let
$\EE_{P_1}$ be the infinite--\DIM al analogue of the open unit
disk~\cite{Si,Se}, consisting of all \HS\ \OP s $Z$ from $\KK_1$ to
$\KK_2$ which are {\em\SYM ic} in the sense that
\begin{equation}
  \label{SYM}
  Z=\4{Z^*}  
\end{equation}
and have norm less than 1 (the latter condition is equivalent to
$P_1+Z\+Z$ being positive definite on $\KK_1$, since $Z\+=-Z^*$ and $Z$
is compact). Then the following is more or less well--known
(cf.~\cite{Se}).
\begin{Prop}\label{prop:BP}
  $P\mapsto P_{21}{P_{11}}^{-1}$ defines a bijection from $\PP_{P_1}$
  onto $\EE_{P_1}$, with inverse given by
  \begin{equation} \label{PZ}
    Z\mapsto P_Z\DEF(P_1+Z)(P_1+Z\+Z)^{-1}(P_1+Z\+).
  \end{equation}
  The restricted symplectic group $\SP{P_1}{0}$ acts transitively on
  either set, in a way compatible with the above bijection, through
  the formulas
  \begin{align}
    P&\mapsto UPU\+\label{OP1}\\
    Z&\mapsto(U_{21}+U_{22}Z)(U_{11}+U_{12}Z)^{-1}. 
      \tag{\ref{OP1}$'$}\label{OP2}
  \end{align}
  The restrictions of these actions to the subgroup $\SP{\text{\rm
      HS}}{}$ remain transitive, as follows from the fact that, for
  $Z\in\EE_{P_1}$,
  \begin{equation}
    \label{UZ}
    U_Z\DEF(P_1+Z)(P_1+Z\+Z)^{-\2}+(P_2-Z\+)(P_2+ZZ\+)^{-\2}    
  \end{equation}
  lies in $\SP{\text{\rm HS}}{}$ and fulfills $U_ZP_1U_Z\+=P_Z$
  (equivalently, under the action \eqref{OP2}, $U_Z$ takes
  $0\in\EE_{P_1}$ to $Z$).
\end{Prop}
\begin{proof}
  Having made $\KK$ into a \HSP, the conditions on $P$ to be a
  \BP~\eqref{BP} may be rewritten as
  \begin{equation}
    \label{BP0}
    P=P\+=\1-\4{P}=P^2,\qquad CP\text{ is positive definite on }\RAN P;  
  \end{equation}
  or, in components:
  \begin{subequations}
    \begin{alignat}{2}
      P_{11}&={P_{11}}^*&&=P_1-\4{P_{22}},\label{BP1}\\
      P_{22}&={P_{22}}^*&&=P_2-\4{P_{11}},\label{BP2}\\
      P_{21}&=\4{{P_{21}}^*}&&=-{P_{12}}^*,\label{BP3}
    \end{alignat}
    \begin{align}
      {P_{11}}^2-P_{11}&={P_{21}}^*P_{21},\label{BP4}\\
      {P_{22}}^2-P_{22}&={P_{12}}^*P_{12},\label{BP5}\\
      (P_1-P_{11})P_{12}&=P_{12}P_{22},\label{BP6}\\
      (P_2-P_{22})P_{21}&=P_{21}P_{11},\label{BP7}
    \end{align}
    \begin{equation}
      \label{BP8}
      \begin{pmatrix}
        P_{11} & P_{12} \\ -P_{21} & -P_{22}
      \end{pmatrix}
      \text{ is positive definite on\,}\RAN P.
    \end{equation}
  \end{subequations}
  Moreover, $P_1-P$ is \HS\ \IFF $P_2P$ is.

  Now let $P\in\PP_{P_1}$. Then $P_{22}\leq0$ by \eqref{BP8},
  hence, by \eqref{BP1}, 
  $$P_{11}=P_1-\4{P_{22}}\geq P_1$$ has a bounded inverse. Thus $Z\DEF
  P_{21}{P_{11}}^{-1}$ is a well--defined \HS\ \OP. By
  \eqref{BP1}--\eqref{BP3} and \eqref{BP7},
  \begin{equation*}
    \begin{split}
      Z-\4{Z^*}&=P_{21}{P_{11}}^{-1}-\4{{P_{11}}^{-1}{P_{21}}^*}\\
      &=\4{{P_{11}}^{-1}}\bigl((P_2-P_{22})P_{21}
      -P_{21}P_{11}\bigr){P_{11}}^{-1}\\
      &=0,
    \end{split} 
  \end{equation*}
  so $Z$ is \SYM ic in the sense of \eqref{SYM}. Furthermore,
  by~\eqref{BP4},
  \begin{equation}
    \label{Z}
    \begin{split}
      P_1-Z^*Z&=P_1-{P_{11}}^{-1}{P_{21}}^*P_{21}{P_{11}}^{-1}\\
      &=P_1-{P_{11}}^{-1}({P_{11}}^2-P_{11}){P_{11}}^{-1}\\
      &={P_{11}}^{-1}
    \end{split}
  \end{equation}
  is positive definite on $\KK_1$, which proves $Z\in\EE_{P_1}$.

  Next let $Z\in\EE_{P_1}$ and let $P_Z$ be given by \eqref{PZ}. We
  \ASS e with $Z$ an \OP 
  \begin{equation}
    \label{Y}
    Y\DEF(P_1+Z\+Z)^{-1}=(P_1-Z^*Z)^{-1}
  \end{equation}
  which is bounded by assumption. Then $P_Z=P_Z\+=P_Z^2$ since
  $(P_1+Z\+)(P_1+Z)=Y^{-1}$.  To prove that $P_Z+\4{P_Z}=\1$ holds,
  note that $ZY^{-1}=\4{Y}^{\,-1}Z$ and therefore $\4{Y}Z=ZY$,
  $YZ\+=Z\+\4{Y}$. It follows that
  \begin{equation*}
    \begin{split}
      P_Z+\4{P_Z}&=(P_1+Z)Y(P_1+Z\+)+(P_2-Z\+)\4{Y}(P_2-Z)\\
      &=Y+ZY+YZ\++ZZ\+\4{Y}+\4{Y}-YZ\+-ZY+Z\+ZY\\
      &=Y^{-1}Y+\4{Y}^{\,-1}\4{Y}\\
      &=P_1+P_2\\
      &=\1. 
    \end{split} 
  \end{equation*}
  Since $P_2P_Z$ is clearly HS and since 
  \begin{equation}
    \label{CPZ}
    CP_Z=(P_1-Z)Y(P_1-Z^*)    
  \end{equation}
  is positive definite on $\RAN P_Z=\RAN(P_1+Z)$, we get that
  $P_Z\in\PP_{P_1}$ as desired.

  To show that these two maps are mutually inverse, let first
  $Z\in\EE_{P_1}$. 
  Then $(P_Z)_{21}{(P_Z)_{11}}^{-1}=ZYY^{-1}=Z$. Conversely, let
  $P\in\PP_{P_1}$ be given and set $Z\DEF P_{21}{P_{11}}^{-1}$. Then
  $ZP_{11}=P_{21}$ and $P_{11}Z\+={P_{21}}\+=P_{12}$. By \eqref{Z} and
  \eqref{Y}, $Y=P_{11}$, hence $P_{11}Z\+=Z\+\4{P_{11}}$. Thus we get 
  \begin{equation*}
    \begin{split}
      P-P_Z&=P-(P_1+Z)P_{11}(P_1+Z\+)\\
      &=P-P_{11}-ZP_{11}-P_{11}Z\+-ZP_{11}Z\+\\
      &=P-P_{11}-P_{21}-P_{12}-ZZ\+\4{P_{11}}\\
      &=P_{22}-ZZ\+\4{P_{11}}\\
      &=P_2-(P_2+ZZ\+)\4{P_{11}}\text{ (by \eqref{BP2})}\\
      &=0.
    \end{split} 
  \end{equation*}
  
  It remains to prove the statements about the group actions. It is
  fairly obvious that $\SP{P_1}{0}$ acts on $\PP_{P_1}$ via
  \eqref{OP1}. The proof that $U_Z$ is a \BOG\OP\ which takes $P_1$ to
  $P_Z$ is also straightforward. To show that
  $U_Z\in\SP{\text{HS}}{}$, let $Y$ be given by \eqref{Y}. Then
  $$Y^\2-P_1=Y^\2(P_1-Y^{-1})(P_1+Y^{-\2})^{-1}
  =Y^\2Z^*Z(P_1+Y^{-\2})^{-1}$$ is of trace class. Therefore
  $(U_Z-\1)P_1=(P_1+Z)Y^\2-P_1=Y^\2-P_1+ZY^\2$ is HS, which implies
  $U_Z\in\SP{\text{HS}}{}$.

  Finally we have to show that the action \eqref{OP1} on $\PP_{P_1}$
  carries over to an action \eqref{OP2} on $\EE_{P_1}$. Thus, for
  given $Z\in\EE_{P_1}$ and $U\in\SP{P_1}{0}$, we have to compute the
  \OP\ $Z'=P'_{21}{P'_{11}}^{-1}$ which \COR s to $P'=UP_ZU\+$. By
  definition, 
  \begin{equation}
    \label{P'}
    \begin{aligned}
      P'_{21}&=(U_{21}+U_{22}Z)Y(U_{11}+U_{12}Z)^*,\\
      P'_{11}&=(U_{11}+U_{12}Z)Y(U_{11}+U_{12}Z)^*.
    \end{aligned}
  \end{equation}
  Suppose that $(U_{11}+U_{12}Z)f=0$ for some $f\in\KK_1$. Then
  $\NORM{f}=\NORM{{U_{11}}^{-1}U_{12}Zf}$. Since
  $\NORM{{U_{11}}^{-1}U_{12}}^2 
  =\NORM{{U_{12}}^*{U_{11}}^{-1*} {U_{11}}^{-1}U_{12}}
  =\NORM{{U_{12}}^*(P_1+U_{12}{U_{12}}^*)^{-1}U_{12}}
  =\NORM{U_{12}}^2/(1+\NORM{U_{12}}^2)<1$
  and $\NORM{Z}<1$, it follows that $f=0$. Hence $U_{11}+U_{12}Z$ is
  injective, and, as a Fredholm \OP\ with vanishing index \eqref{IND},
  it has a bounded inverse. So we get from \eqref{P'} that
  $Z'=P'_{21}{P'_{11}}^{-1}=(U_{21}+U_{22}Z)(U_{11}+U_{12}Z)^{-1}$ as
  claimed.
\end{proof}
The following construction will enable us to assign, in an unambiguous
way, to each \BOG\OP\ $V\in\SP{P_1}{}$ a \BP\ $P_V$ such that
\eqref{PV} holds.
\begin{Lem}
  \label{lem:H}
  Let $\HH\subset\KK$ be a closed *--invariant subspace such that
  $\gamma|_{\HH\times \HH}$ is nondegenerate and such that $[P_1,E]$
  is \HS\ where $E$ is the orthogonal projection onto $\HH$. Let
  $A\DEF ECE$ be the self--adjoint \OP, invertible on $\HH$, such that
  $\gamma(f,g)=\langle{f,Ag}\rangle,\ f,g\in\HH$, and let $A_\pm$ be
  the unique positive \OP s such that $A=A_+-A_-$ and $A_+A_-=0$.
  Further let $A^{-1}$ be defined as the inverse of $A$ on $\HH$ and
  as zero on $\HH^\bot$, and similarly for $A_\pm^{-1}$. Then
  $A^{-1}C$ is the $\gamma$--orthogonal projection onto $\HH$,
  $P_+\DEF A_+^{-1}C$ is a \BP\ of $\HH$, and $P_2P_+$ is \HS.
  Moreover, $P_+=P_1E$ \IFF $[P_1,E]=0$.
\end{Lem}
\begin{proof}
  Let $E'\DEF \1-E$. Since $ECE'$ and $E'CE$ are compact by
  assumption, $C-ECE'-E'CE=A+E'CE'$ is a Fredholm \OP\ on $\KK$ with
  vanishing index. Hence $A$ is Fredholm on $\HH$ with $\IND A=0$. $A$
  is injective since $\gamma$ is nondegenerate on $\HH$. It is
  therefore a bounded bijection on $\HH$ with a bounded inverse (the
  same holds true for $A_\pm$ as \OP s on $\RAN A_\pm$). Thus $Q\DEF
  A^{-1}C$ is well--defined. It fulfills $Q^2=A^{-1}(ECE)A^{-1}C=Q$
  and $Q\+=C(CA^{-1})C=Q$. So $Q$ is a projection, self--adjoint \WRT
  $\gamma$. Since its range equals $\RAN A^{-1}=\HH$, it is the
  $\gamma$--orthogonal projection onto $\HH$.

  By a similar argument, $P_+$ is also a $\gamma$--orthogonal
  projection.  It is straightforward to see that $P_+=P_1E$ \IFF
  $[P_1,E]=0$. To show that $P_+$ is actually a \BP\ of $\HH$
  (cf.~\eqref{BP0}), note that $\4{A_+}=A_-$ because of $\4{A}=-A$
  (and uniqueness of $A_\pm$). This implies
  $P_++\4{P_+}=A_+^{-1}C-A_-^{-1}C=A^{-1}C=\1_\HH$. Positive
  definiteness of $CP_+$ on $\RAN P_+$ follows from
  $\langle{f,CP_+f}\rangle=\NORM{A_+^{-1/2}Cf}^2$.

  To prove that $P_2P_+$ is HS, let $D\DEF EP_1E-A_+$. Since
  $EP_1E-EP_2E=A=A_+-A_-$, we have $D=\4{D}$. We claim that $D$ is of
  trace class. Since $ECE'$ is HS,
  \begin{eqnarray*}
    ECE'CE &=& EC(\1-E)CE\\
    &=& E-(ECE)^2\\
    &=& E-A^2\\
    &=& (E+\ABS{A})(E-\ABS{A})
  \end{eqnarray*}
  is of trace class. Since $E+\ABS{A}$ has a bounded inverse (as an
  \OP\ on $\HH$) and since $\ABS{A}=A_++A_-$, it follows that
  $E-\ABS{A}=EP_1E+EP_2E-A_+-A_-=D+\4{D}=2D$ is of trace class as
  claimed.  As a consequence, $A_+P_2=(EP_1E-D)P_2$ is HS ($P_1EP_2$
  is HS by assumption). By boundedness of $A_+^{-1}$,
  $P_+P_2=-A_+^{-2}(A_+P_2)$ and $P_2P_+=(P_+P_2)\+$ are also HS. This
  completes the proof.
\end{proof}
Now let $V\in\SP{P_1}{}$. We already showed in Section~\ref{sec:IMP}
that the restriction of $\gamma$ to $\ker V\+$ is nondegenerate. We
also showed in the proof of Theorem~\ref{th:IMP} that $[P_1,V']$ is
\HS\ where $V'$ is the \ISM y arising from polar \DEC ition of $V$.
Hence $[P_1,E]$ is \HS\ where $E=C(\1-V'{V'}^*)C$ is the orthogonal
projection onto $\ker V\+$. Thus Lemma~\ref{lem:H} applies to
$\HH=\ker V\+$.
\begin{Def}\label{def:UW}
  For $V\in\SP{P_1}{}$, let $P_{V+}$ be the \BP\ of $\ker V\+$ given
  by Lemma~\ref{lem:H}, and set
  \begin{alignat*}{2}
    P_V&\DEF VP_1V\++P_{V+}&&\in\PP_{P_1},\\
    Z_V&\DEF(P_V)_{21}{(P_V)_{11}}^{-1}&&\in\EE_{P_1}
  \end{alignat*}
  (cf.\ Proposition~\ref{prop:BP}). Further let $U_V\in\SP{\text{\rm
      HS}}{}$ be the \BOG\OP\ \ASS ed with $Z_V$ according to \eqref{UZ},
  and define $W_V\DEF U_V\+V\in\SP{\text{\rm diag}}{}$.
\end{Def}\noindent
$P_V$ clearly is a \BP\ which satisfies \eqref{PV}. Actually, any \BP\ 
$P$ fulfilling $V\+PV=P_1$ or, equivalently, $PV=VP_1$, is of the
form $P=VP_1V\++P'$ where $P'$ is some \BP\ of $\ker V\+$. What had to
be proved above is that $P'$ can be chosen such that $P_2P'$ is \HS,
in the case $\dim\ker V\+=\infty$. In fact, any such choice would
suffice for what follows.

The condition $V\+P_VV=P_1$ translates into the condition 
\begin{equation}
  \label{ZV}
  Z_VV_{11}=V_{21}
\end{equation}
for $Z_V$. Again, each $Z\in\EE_{P_1}$ fulfilling \eqref{ZV} would do,
but we prefer to have a definite choice. It follows from \SYM y
\eqref{SYM} that any $Z$ which solves \eqref{ZV} must have the form
\begin{equation}
  \label{ZV1}
  Z=V_{21}{V_{11}}^{-1}+{V_{22}}^{-1*}{V_{12}}^*P_{\ker{V_{11}}^*}+Z'
\end{equation}
where $P_\HH$ denotes the orthogonal projection onto some closed
subspace $\HH\subset\KK$, ${V_{11}}^{-1}$ and ${V_{22}}^{-1}$ have
been defined below \eqref{REL}, and $Z'$ is a \SYM ic \HS\ \OP\ from
$\ker{V_{11}}^*$ to $\ker{V_{22}}^*$. The freedom in the choice of
$Z'$ \COR s to the freedom in the choice of $P'$. Note that $Z$ can be
written, \WRT the \DEC itions $\KK_1=\RAN V_{11}\oplus\ker{V_{11}}^*,\ 
\KK_2=\RAN V_{22}\oplus\ker{V_{22}}^*$, as
\begin{equation}
  Z=\begin{pmatrix}
    P_{\RAN{V_{22}}}V_{21}{V_{11}}^{-1} & 
    {V_{22}}^{-1*}{V_{12}}^*P_{\ker{V_{11}}^*} \\
    P_{\ker{V_{22}}^*}V_{21}{V_{11}}^{-1} & Z'
  \end{pmatrix}.\tag{\ref{ZV1}$'$}\label{ZV2}
\end{equation}
The \HS\ norm of $Z$ is minimized by choosing $Z'=0$, but there are
examples in which this choice violates the condition $\NORM{Z}<1$,
i.e.\ it does not always define an element of $\EE_{P_1}$. This is in
contrast to the CAR case where the choice analogous to $Z'=0$ appears
to be natural \cite{CB}. As we shall see in Section~\ref{sec:CON},
$Z_V$ describes the values of \IMP ers on the Fock vacuum.

The \OP s $U_V$ and $W_V$ constitute the product \DEC ition of $V$
that was announced earlier. $W_V$ is diagonal because
$P_1W_V=P_1U_V\+V=U_V\+P_VV=U_V\+VP_1=W_VP_1$. Explicitly, one
computes that
$$W_V=\begin{pmatrix}
  (P_1+Z_V\+Z_V)^\2V_{11} & 0 \\ 0 & (P_2+Z_VZ_V\+)^\2V_{22}
\end{pmatrix}$$
\WRT the \DEC ition $\KK=\KK_1\oplus\KK_2$. Let us summarize the
properties of these \OP s.
\begin{Prop}
  \label{prop:UW}
  Definition \ref{def:UW} establishes a \DEC ition of $V\in\SP{P_1}{}$
  $$V=U_VW_V$$ where $U_V\in\SP{\text{\rm HS}}{}$
  and $W_V\in\SP{\text{\rm diag}}{}$ have the properties
  \begin{align*}
    \IND U_V&=0, & Z_{U_V}&=Z_V, & P_{U_V}&=P_V;\\ 
    \IND W_V&=\IND V, & Z_{W_V}&=0, & P_{W_V}&=P_1.
  \end{align*}
  In particular, if $V\in\SP{P_1}{0}$, then 
  $$U_V=\begin{pmatrix}
    \ABS{{V_{11}}^*} & V_{12}{v_{22}}^* \\ 
    V_{21}{v_{11}}^* & \ABS{{V_{22}}^*}
  \end{pmatrix},\qquad 
  W_V=\begin{pmatrix}
    v_{11} & 0 \\ 0 & v_{22}
  \end{pmatrix}$$ 
  where $v_{11}\DEF V_{11}\ABS{V_{11}}^{-1}$ and $v_{22}=\4{v_{11}}$
  are the unitary parts of $V_{11}$ and $V_{22}$; whereas if
  $V\in\SP{\text{\rm diag}}{}$, then $U_V=\1$ and $W_V=V$.
\end{Prop}
\begin{Rem}
  The product \DEC ition described above is the generalization to the
  infinite--\DIM al case of a construction given by Maa\ss\ \cite{M}.
  The exact analogue of the construction given in \cite{CB} in the
  fermionic case would be to define $W'\in\SP{\text{diag}}{}$ through
  $W'_{11}\DEF V_{11} \ABS{V_{11}}^{-1}$ (the \ISM ic part of
  $V_{11}$), to choose a \BOG\OP\ $u'$ from $\ker {W'}\+$ to $\ker
  V\+$ such that $u'P_1=P_Vu'$, and to set $U'\DEF
  V{W'}\++u'\in\SP{P_1}{0}$. Then $U'$ and $W'$ would also have the
  properties listed in Proposition~\ref{prop:UW}, with the exception
  that $U'-\1$ is not necessarily \HS. On the other hand, this choice
  has the merit that the definition of $W'$ is completely canonical
  (independent of the choice of $Z$). 

  Though it was not shown in \cite{CB}, it holds true also in the CAR
  case that each \IMP able \BOG\OP\ can be written as a product of two
  factors where the first differs from $\1$ only by a \HS\ part, and
  the second is diagonal.
\end{Rem}
\begin{Cor}
  \label{cor:SP}
  $\SP{P_1}{}=\SP{\text{\rm HS}}{}\cdot\SP{\text{\rm diag}}{}$.
  The orbits of the action of $\SP{P_1}{0}$ on $\SP{P_1}{}$ are the
  subsets $\SP{P_1}{n},\ n\in\NN\cup\{\infty\}$. They coincide with
  the connected components of $\SP{P_1}{}$.
\end{Cor}
\section{Normal Form of Cuntz Algebra Generators}
\label{sec:CON}
The first step in the construction of \IMP ers consists in a
generalization of the definition of `bilinear Hamiltonians' \cite{A71}
from the finite rank case to the case of bounded \OP s. If $H$ is a
finite rank \OP\ on $\KK$ such that $H^*=\4{H}=-H$, then $e^{HC}$
belongs to $\SP{\text{HS}}{}$. Expanding $H=\sum f_j\langle
g_j,.\rangle$, one obtains a skew--adjoint element $b_0(H)\DEF\sum
f_jg_j^*$ of $\CK$ which is a linear function of $H$, independent of
the choice of $f_j,g_j\in\KK$. Then $\PIP\bigl(b_0(H)\bigr)$ is
essentially skew--adjoint on $\DD$, and, if $b(H)$ denotes its
closure, $\exp\bigl(\2b(H)\bigr)$ is a unitary which \IMP s the \AUT\ 
induced by $e^{HC}$ \cite{A71,A82}.

Using Wick ordering, the definition of bilinear Hamiltonians can be
extended to arbitrary bounded \OP s $H$ which are \SYM ic in the sense
of \eqref{SYM}\footnote{The bilinear Hamiltonian \COR ing to an
  anti\SYM ic \OP\ ($H=-\4{H^*}$) vanishes.}:
\begin{equation}
  \label{SYMH}
  H_{11}=\4{{H_{22}}^*},\qquad H_{12}=\4{{H_{12}}^*},\qquad
  H_{21}=\4{{H_{21}}^*}.  
\end{equation}
Without loss of generality, we henceforth assume that
$\KK_1=L^2(\RR^d)$. Then let $\SSS\subset\FK$ be the dense subspace
consisting of finite particle vectors $\phi$ with $n$--particle wave
functions $\phi^{(n)}$ in the Schwartz space $\SSS(\RR^{dn})$. The
unsmeared \ANN ion \OP\ $a(p)$ with (invariant) domain $\SSS$ is
defined as usual
$$(a(p)\phi)^{(n)}(p_1,\dots,p_n)\DEF\sqrt{n+1}\,\phi^{(n+1)}
(p,p_1,\dots,p_n).$$ Let $a^*(p)$ be its \QF\ adjoint on
$\SSS\times\SSS$. Then Wick ordered monomials $a^*(q_1)\dotsm
a^*(q_m)a(p_1)\dotsm a(p_n)$ make sense as \QF s on $\SSS\times\SSS$
\cite{GJ,RS}, and, for $\phi,\phi'\in\SSS$, 
$$\langle\phi,a^*(q_1)\dotsm a^*(q_m)a(p_1)\dotsm a(p_n)\phi'\rangle
\DEF\langle a(q_1)\dotsm a(q_m)\phi,a(p_1)\dotsm a(p_n)\phi'\rangle$$
is a Schwartz function to which tempered distributions may be applied.
In particular, the distributions $H_{jk}(p,q),\ j,k=1,2$, given by
\begin{align*}
  \langle{f,H_{11}g}\rangle&=\int\4{f(p)}H_{11}(p,q)g(q)\,dp\,dq,\\
  \langle{f,H_{12}g^*}\rangle&=\int\4{f(p)}H_{12}(p,q)\4{g(q)}\,dp\,dq,\\
  \langle{f^*,H_{21}g}\rangle&=\int f(p)H_{21}(p,q)g(q)\,dp\,dq,\\
  \langle{f^*,H_{22}g^*}\rangle&=\int f(p)H_{22}(p,q)\4{g(q)}\,dp\,dq
\end{align*}
for $f,g\in\SSS(\RR^d)\subset\KK_1$, give rise to the following \QF s
on $\SSS\times\SSS$: 
\begin{align*}
  H_{11}a^*a &\DEF \int a(p)^*H_{11}(p,q)a(q)\,dp\,dq \\
  H_{12}a^*a^* &\DEF \int a(p)^*H_{12}(p,q)a(q)^*\,dp\,dq \\
  H_{21}aa &\DEF \int a(p)H_{21}(p,q)a(q)\,dp\,dq\\
  \WO{H_{22}aa^*} &\DEF\int a(q)^*H_{22}(p,q)a(p)\,dp\,dq=H_{11}a^*a.
\end{align*}
Wick ordering of $H_{22}aa^*$ is necessary to make this expression
well--defined. The last equality follows from \SYM y of $H$:
$$H_{11}(p,q)=H_{22}(q,p),\quad H_{12}(p,q)=H_{12}(q,p),\quad
H_{21}(p,q)=H_{21}(q,p).$$ 
We next define $\WO{b(H)}$ and its Wick ordered powers as \QF s on
$\SSS\times\SSS$:
\begin{align*}
  \WO{b(H)}\DEF&\ H_{12}a^*a^*+2H_{11}a^*a+H_{21}aa,\\
  \WO{b(H)^l}\DEF&\ l!\sum_{\substack{l_1,l_2,l_3=0 \\ l_1+l_2+l_3=l}}^l
    \frac{2^{l_2}}{l_1!l_2!l_3!}H_{l_1,l_2,l_3},\qquad l\in\NN,\\ 
  \text{with }H_{l_1,l_2,l_3} \DEF&\ 
    \int H_{12}(p_1,q_1)\dotsm H_{12}(p_{l_1},q_{l_1})
    H_{11}(p_1',q_1')\dotsm H_{11}(p_{l_2}',q_{l_2}')\\
  &{\ }\cdot H_{21}(p_1'',q_1'')\dotsm H_{21}(p_{l_3}'',q_{l_3}'')
    a^*(p_1)\dotsm a^*(p_{l_1})a^*(q_1)\dotsm a^*(q_{l_1})\\
  &{\ }\cdot a^*(p_1')\dotsm a^*(p_{l_2}') a(q_1')\dotsm a(q_{l_2}')
    a(p_1'')\dotsm a(p_{l_3}'')a(q_1'')\dotsm a(q_{l_3}'')\\
  &{\ }\cdot dp_1\,dq_1\dots dp_{l_1}\,dq_{l_1}\,dp_1'\,dq_1'\dots
    dp_{l_2}'\,dq_{l_2}'\,dp_1''\,dq_1''\dots dp_{l_3}''\,dq_{l_3}''.
\end{align*}
The Wick ordered exponential of $\2b(H)$ is also well-defined on
$\SSS\times\SSS$, since only a finite number of terms contributes when
applied to vectors from $\SSS$:
$$\WO{\exp\left(\2b(H)\right)}\DEF
\sum_{l=0}^\infty\frac{1}{l!2^l}\WO{b(H)^l}.$$ 
The important point is that these \QF s are actually the forms of
uniquely determined linear \OP s, defined on the dense subspace $\DD$
and mapping $\DD$ into the domain of (the closure of) any creation or
\ANN ion \OP, provided that \cite{R78}
\begin{equation}
  \label{OPH}
  \NORM{H_{12}}<1,\qquad H_{12}\text{ is \HS.}
\end{equation}
These \OP s will be denoted by the same symbols as the \QF s.
\begin{Lem}\label{lem:CR}
  Let $H\in\BB(\KK)$ satisfy \eqref{SYMH} and \eqref{OPH}. Then the
  following commutation relations hold on $\DD$, for $f\in\KK_1$:
  \begin{align*}
    [H_{l_1,l_2,l_3},a(f)^*] &=l_2a(H_{11}f)^*H_{l_1,l_2-1,l_3}
      +2l_3H_{l_1,l_2,l_3-1}a\bigl((H_{21}f)^*\bigr),\\
    [a(f),H_{l_1,l_2,l_3}] &=2l_1a(H_{12}f^*)^*H_{l_1-1,l_2,l_3}+
      l_2H_{l_1,l_2-1,l_3}a({H_{11}}^*f),
  \end{align*}
  implying that
  \begin{multline*}
    \left[\WO{\exp\left(\2b(H)\right)},a(f)^*\right] \\
    \shoveright{=a(H_{11}f)^*\WO{\exp\left(\2b(H)\right)}
      +\WO{\exp\left(\2b(H)\right)}a\bigl((H_{21}f)^*\bigr),}\\
    \shoveleft{\left[a(f),\WO{\exp\left(\2b(H)\right)}\right]} \\
    {=a(H_{12}f^*)^*\WO{\exp\left(\2b(H)\right)}
      +\WO{\exp\left(\2b(H)\right)}a({H_{11}}^*f).}
  \end{multline*}
\end{Lem}
\begin{proof}
  Compute as in \cite{R78,CB}.
\end{proof}
For given $V\in\SP{P_1}{}$, we are now looking for \SYM ic bounded \OP
s $H$ which satisfy \eqref{OPH} and the following intertwiner relation
on $\DD$
\begin{equation}
  \label{INT}
  \WO{\exp\left(\2b(H)\right)}\PIP(f)=
  \PIP(Vf)\WO{\exp\left(\2b(H)\right)},\quad f\in\KK
\end{equation}
(taking the closure of $\PIP(Vf)$ is tacitly assumed here). This
problem turns out to be equivalent to the determination of the \OP s
$Z$ done in \eqref{ZV}, \eqref{ZV1}.
\begin{Lem}\label{lem:HZ}
  Each $Z\in\EE_{P_1}$ fulfilling \eqref{ZV} gives rise to a unique
  solution $H$ of the above problem through the formula
  $$H=\begin{pmatrix}
    V_{11}-P_1+Z\+V_{21} & Z\+ \\
    ({V_{22}}^*+{V_{12}}^*Z\+)V_{21} & {V_{22}}^*-P_2+{V_{12}}^*Z\+
  \end{pmatrix},$$ 
  and each solution arises in this way.
\end{Lem}
\begin{proof}
  Let us abbreviate $\eta_H\DEF\WO{\exp(\2b(H))}$. Choosing $f\in\KK_2$
  resp.\ $f\in\KK_1$ and inserting the definition of $\PIP$, one finds
  that \eqref{INT} is equivalent to 
  $$\eta_H a(g)=\bigl(a(V_{11}g)+a^*(V_{12}g^*)\bigr)\eta_H,\quad \eta_H
  a^*(g)=\bigl(a^*(V_{11}g)+a(V_{12}g^*)\bigr)\eta_H$$ for $g\in\KK_1$.
  Using the commutation relations from Lemma~\ref{lem:CR}, these
  equations may be brought into Wick ordered form:
  \begin{align*}
    0 &= a^*\bigl((V_{12}+H_{12}V_{22})g^*\bigr)\eta_H + 
      \eta_H a\Bigl(\bigl((P_1+{H_{11}}^*)V_{11}-P_1\bigr)g\Bigr),\\
    0 &= a^*\bigl((P_1+H_{11}-V_{11}-H_{12}V_{21})g\bigr)\eta_H +
      \eta_H a\Bigl(\bigl(\4{H_{21}}-(P_1+{H_{11}}^*)V_{12}\bigr)g^*\Bigr).
  \end{align*}
  As in the CAR case \cite{CB}, these equations hold for all
  $g\in\KK_1$ \IFF 
  \begin{subequations}
    \begin{align}
      0&= V_{12}+H_{12}V_{22},\label{H1}\\
      0&= P_1+H_{11}-V_{11}-H_{12}V_{21},\label{H2}\\
      0&= H_{21}-(P_2+H_{22})V_{21},\label{H3}\\
      0&= P_2-(P_2+H_{22})V_{22}\label{H4}
    \end{align}
  \end{subequations}
  (we applied complex conjugation and used $\4{{H_{11}}^*}=H_{22}$).

  Now assume that $H$ solves the above problem. It is then obvious
  from \eqref{SYMH}, \eqref{OPH} and \eqref{H1} that $Z\DEF{H_{12}}\+$
  belongs to $\EE_{P_1}$ and fulfills \eqref{ZV}.

  Conversely, let $Z\in\EE_{P_1}$ satisfy \eqref{ZV}. If there exists
  a solution $H$ with $H_{12}=Z\+$, then $H_{11}$ is fixed by
  \eqref{H2}, $H_{22}$ must equal $\4{{H_{11}}^*}$, and $H_{21}$ is
  determined by \eqref{H3}. Thus there can be at most one solution
  \COR ing to $Z$, and it is necessarily of the form stated in the
  proposition.

  It remains to prove that the so--defined $H$ has all desired
  properties, i.e.\ that $H_{21}$ is \SYM ic and that \eqref{H4}
  holds, the rest being clear by construction. The first claim follows
  from \eqref{REL4}:
  $$H_{21}-\4{{H_{21}}^*}=({V_{22}}^*+{V_{12}}^*Z\+)V_{21}-
  {V_{12}}^*(V_{11}+Z\+V_{21})=0,$$ and the second from \eqref{ZV} and
  \eqref{REL2}: 
  $$(P_2+H_{22})V_{22}=({V_{22}}^*-{V_{12}}^*\4{Z})V_{22}=
  {V_{22}}^*V_{22}-{V_{12}}^*V_{12}=P_2.$$ 
\end{proof}
Inserting the formula~\eqref{ZV1} for $Z$, one obtains 
\begin{eqnarray*}
  H_{11} &=& {V_{11}}^{-1*}-P_1-P_{{\ker V_{11}}^*}
             V_{12}{V_{22}}^{-1}V_{21} + {Z'}\+V_{21},\\
  H_{12} &=& -V_{12}{V_{22}}^{-1}-{V_{11}}^{-1*}{V_{21}}^* 
             P_{{\ker V_{22}}^*} + {Z'}\+,\\
  H_{21} &=& ({V_{22}}^{-1}-{V_{12}}^*{V_{11}}^{-1*}{V_{21}}^*
             P_{{\ker V_{22}}^*})V_{21} + {V_{12}}^*{Z'}\+V_{21},\\
  H_{22} &=& {V_{22}}^{-1}-P_2-{V_{12}}^*{V_{11}}^{-1*}{V_{21}}^*
             P_{{\ker V_{22}}^*} + {V_{12}}^*{Z'}\+.
\end{eqnarray*}
$H$ \COR s to \RUI' \OP\ $\Lambda$ \cite{R78}. If one compares the
above formula for $H$ with \RUI' formula for $\Lambda$ in the case of
\AUT s ($\ker {V_{jj}}^*=\{0\},\ j=1,2,\ Z'=0$), one finds that the
off--diagonal components carry opposite signs. This is due to the fact
that \RUI\ actually constructs \IMP ers for the transformation induced
by $CVC$ rather than $V$, cf.\ (3.27) and (3.29) in \cite{R78}.

Note that $\WO{\exp\bigl(\2b(H)\bigr)}\Omega_{P_1}
=\exp(\2H_{12}a^*a^*)\Omega_{P_1}$. By \RUI' computation \cite{R78}
(see also \cite{Se}), the norm of such vectors is
$$\left\|\WO{\exp\left(\2b(H)\right)}\Omega_{P_1}\right\|=
\left(\det(P_1+H_{12}{H_{12}}\+)\right)^{-1/4}.$$
\begin{Def}
  \label{def:PSIA}
  Let $V\in\SP{P_1}{}$, and let $P_V,\ Z_V$ and $H_V$ be the \OP s \ASS
  ed with $V$ according to Definition~\ref{def:UW} and
  Lemma~\ref{lem:HZ}. Choose a $\gamma$--\ONB\ $f_1,f_2,\dotsc$ in
  $P_V(\ker V\+)$, i.e.\ a basis such that
  $\gamma(f_j,f_k)=\delta_{jk}$ (this is possible because the
  restriction of $\gamma$ to $P_V(\ker V\+)$ is positive definite).
  Let $\psi_j$ be the \ISM y obtained by polar \DEC ition of the
  closure of $\PIP(f_j)$. Then define \OP s $\Psi_\alpha(V)$ on $\DD$,
  for any multi--index $\alpha=(\alpha_1,\dots,\alpha_l)$ with
  $\alpha_j\leq\alpha_{j+1}$ (or $\alpha=0$) as in \eqref{ALPHA}, as
  \begin{equation}
    \label{PSIA}
    \Psi_\alpha(V)\DEF\left(\det(P_1+Z_V\+Z_V)\right)^{\frac{1}{4}}
    \psi_{\alpha_1}\dotsm\psi_{\alpha_l}\WO{\exp\left(\2b(H_V)\right)}.
  \end{equation}
\end{Def}
\begin{Th}
  \label{th:PSIA}
  The $\Psi_\alpha(V)$ extend continuously to \ISM ies (denoted by the
  same symbols) on the \SYM ic Fock space $\FK$ such that
  \begin{equation}
    \label{CUNTZ}
    \Psi_\alpha(V)^*\Psi_\beta(V)=\delta_{\alpha\beta}\1,\quad
    \sum\limits_\alpha\Psi_\alpha(V)\Psi_\alpha(V)^*=\1
  \end{equation}
  and, for any element $w$ of the Weyl algebra $\WK$,
  \begin{equation}
    \label{IMP}
    \R{V}(w)=\sum\limits_\alpha\Psi_\alpha(V)w\Psi_\alpha(V)^*.    
  \end{equation}
\end{Th}
\begin{proof}
  By \eqref{CCR} we have $\PIP(f_j)^*\PIP(f_j)=\1+\PIP(f_j)\PIP(f_j)^*$
  on $\DD$, so the closure of $\PIP(f_j)$ is injective, and $\psi_j$
  is \ISM ic. It is also easy to see, using \eqref{INT}, the CCR and
  $\NORM{\Psi_\alpha(V)\Omega_{P_1}}=1$, that
  \begin{multline*}
    \langle\Psi_\alpha(V)\PIP(g_1\dotsm g_m)\Omega_{P_1},
      \Psi_\alpha(V)\PIP(h_1\dotsm h_n)\Omega_{P_1}\rangle\\
    =\langle\PIP(g_1\dotsm g_m)\Omega_{P_1},
      \PIP(h_1\dotsm h_n)\Omega_{P_1}\rangle.
  \end{multline*}
  Hence $\Psi_\alpha(V)$ is \ISM ic on $\DD$ and has a continuous
  extension to an \ISM y on $\FK$.  

  Let $\HH_j\DEF{\rm{span}}(f_j,f_j^*)$, so that
  $\psi_j\in\WW(\HH_j)''$ by virtue of Lemma~\ref{lem:AFF}. Since
  $\HH_j\subset\ker V\+$, there holds $\WW(\HH_j)\subset\WW(\RAN V)'$
  by duality~\eqref{DUAL}. Now let $f\in\RE\KK$ and $\phi\in\DD$. Since
  $\phi$ is an entire analytic vector for $\PIP(f)$ \cite{A71}, since
  $\DD$ is invariant under $\PIP(f)$, and since $\4{\PIP(Vf)}$ is
  affiliated with $\WW(\RAN V)$ by Lemma~\ref{lem:AFF} (the bar
  denotes closure), it follows from \eqref{INT} that
  \begin{eqnarray*}
    \Psi_\alpha(V)w(f)\phi &=&
    \sum_{n=0}^\infty\frac{i^n}{n!}\Psi_\alpha(V)(\PIP(f))^n\phi\\ 
    &=&\sum_{n=0}^\infty\frac{i^n}{n!}\psi_{\alpha_1}\dotsm
      \psi_{\alpha_l}\bigl(\4{\PIP(Vf)}\bigr)^n\Psi_0(V)\phi\\ 
    &=&\sum_{n=0}^\infty\frac{i^n}{n!}\bigl(\4{\PIP(Vf)}\bigr)^n
      \Psi_\alpha(V)\phi\\  
    &=&w(Vf)\Psi_\alpha(V)\phi.
  \end{eqnarray*}
  By continuity, this entails
  \begin{equation}
    \label{INTA}
    \Psi_\alpha(V)w=\R{V}(w)\Psi_\alpha(V),\qquad w\in\WK.    
  \end{equation}
  We next claim that
  \begin{equation}
    \label{PSI0}
    \psi_j^*\Psi_0(V)=0
  \end{equation}
  or, equivalently, that $\PIP(f_j)^*\Psi_0(V)=0$. To see this, apply
  Lemma~\ref{lem:CR} and write $\PIP(f_j)^*\Psi_0(V)$ in Wick ordered
  form:
  $$\PIP(f_j)^*\Psi_0(V)=a\bigl((P_1+H_{12})f_j^*\bigr)^*\Psi_0(V)+
  \Psi_0(V)a\bigl((P_1+{H_{11}}^*)f_j\bigr)$$
  on $\DD$, with $H\DEF H_V$. Then \eqref{PSI0} holds \IFF
  \begin{equation}
    (P_1+H_{12})f_j^*=0,\qquad(P_1+{H_{11}}^*)f_j=0.
    \tag{\ref{PSI0}$'$}\label{PSI1}
  \end{equation}
  Now $f_j\in\RAN P_V$ is equivalent to $f_j^*\in\ker P_V=\ker
  CP_V=\ker(P_1+H_{12})$ (we used \eqref{CPZ}). This proves the first
  equation in \eqref{PSI1}. It also shows that ${H_{12}}^*f_j=-f_j$.
  Hence by Lemma~\ref{lem:HZ}, 
  $$(P_1+{H_{11}}^*)f_j=({V_{11}}^*+{V_{21}}^*{H_{12}}^*)f_j
  =({V_{11}}^*-{V_{21}}^*)f_j=P_1V\+f_j=0$$ which proves the second
  equation in \eqref{PSI1} and therefore \eqref{PSI0}.

  The orthogonality relation $\Psi_\alpha(V)^*\Psi_\beta(V)=0$
  ($\alpha\not=\beta$) now follows from \eqref{PSI0} and from
  $\WW(\HH_j)\subset\WW(\HH_k)'$ ($j\not=k$) which in turn is a
  consequence of $\gamma(\HH_j,\HH_k)=0$ and \eqref{DUAL}.

  The proof of the completeness relation $\sum\Psi_\alpha(V)
  \Psi_\alpha(V)^*=\1$ is facilitated by invoking the product \DEC
  ition $V=U_VW_V$ from Proposition~\ref{prop:UW}. Set $e_j\DEF
  U_V\+f_j$ to obtain a $\gamma$--\ONB\ $e_1,e_2,\dotsc$ in $P_1(\ker
  W_V\+)=\KK_1\cap\ker W_V\+$. Let $\psi_j'$ be the \ISM ic part of
  $a(e_j)^*$. An application of Definition~\ref{def:PSIA} to $W_V$
  yields \IMP ers $\Psi_\alpha(W_V)=
  \psi'_{\alpha_1}\dotsm\psi'_{\alpha_l}\Psi_0(W_V)$ for $W_V$.
  $Z_{W_V}=0$ entails that
  $$\Psi_\alpha(W_V)\Omega_{P_1}=\psi'_{\alpha_1}\dotsm
  \psi'_{\alpha_l}\Omega_{P_1}.$$ One computes, using the CCR, that
  $\psi'_{\alpha_1}\dotsm \psi'_{\alpha_l}\Omega_{P_1}=\phi'_\alpha$,
  where the $\phi'_\alpha$ are the cyclic vectors \ASS ed with the
  pure state $\omega_{P_1}\0\R{W_V}=\omega_{P_1}$ as in
  Lemma~\ref{lem:CYC}. Let $\FF'_\alpha$ be the closure of $\WW(\RAN
  W_V)\phi'_\alpha$. Since the $\FF'_\alpha$ are irreducible subspaces
  for $\WW(\RAN W_V)$ by Lemma~\ref{lem:CYC}, they must coincide with
  the irreducible subspaces $\RAN\Psi_\alpha(W_V)$.
  $\oplus\FF'_\alpha=\FK$ then implies completeness of the
  $\Psi_\alpha(W_V)$.
 
  The proof will be completed by showing that 
  \begin{equation}
    \label{COMP}
    \Psi_\alpha(V)=\Psi(U_V)\Psi_\alpha(W_V)
  \end{equation}
  holds where $\Psi(U_V)$ is the unitary \IMP er for $U_V$ given by
  Definition~\ref{def:PSIA}. It suffices to show that \eqref{COMP}
  holds on $\Omega_{P_1}$ since any bounded \OP\ fulfilling
  \eqref{INTA} is already determined by its value on $\Omega_{P_1}$.
  Because of $Z_{U_V}=Z_V$ we have
  \begin{equation}
    \label{PSIU}
    \Psi_0(V)\Omega_{P_1}=\Psi(U_V)\Omega_{P_1},    
  \end{equation}
  so it remains to show that
  $\psi_{\alpha_1}\dotsm\psi_{\alpha_l}\Psi(U_V)\Omega_{P_1}=
  \Psi(U_V)\psi'_{\alpha_1}\dotsm\psi'_{\alpha_l}\Omega_{P_1}$. We
  claim that
  $$\psi_j\Psi(U_V)=\Psi(U_V)\psi'_j.$$ For let $T$ (resp.\ $T'$) be
  the closure of $\PIP(f_j)$ (resp.\ $\PIP(e_j)$), and let $T^\pm$
  (resp.\ ${T'}^\pm$) be the \COR ing self--adjoint \OP s as in the
  proof of Lemma~\ref{lem:AFF}, so that 
  $$D(T)=D(T^+)\cap D(T^-),\qquad T=T^+-iT^-,$$ and similar for
  $T'$. Then there holds
  $$\Psi(U_V)\exp(it{T'}^\pm)\Psi(U_V)^*=\exp(itT^\pm),\qquad
  t\in\RR.$$ Therefore $\Psi(U_V)$ maps $D({T'}^\pm)$ onto $D(T^\pm)$,
  and one has $\Psi(U_V){T'}^\pm\Psi(U_V)^*=T^\pm$. Consequently,
  $\Psi(U_V)\bigl(D(T')\bigr)=D(T)$ and $\Psi(U_V)T'\Psi(U_V)^*=T$.
  This implies that $\Psi(U_V)\psi'_j\Psi(U_V)^*=\psi_j$ as claimed.
  The proof is complete since \eqref{CUNTZ} and \eqref{INTA} together
  imply \eqref{IMP}.
\end{proof}
\begin{Cor}
  There is a unitary \ISO ism from $H(\R{V})$, the \HSP\ generated by
  the $\Psi_\alpha(V)$, onto the \SYM ic Fock space $\FF_s(P_V(\ker
  V\+))$ over $P_V(\ker V\+)$, which maps $\Psi_\alpha(V)$ to
  $(l_1!\dotsm l_r!)^{-\2}a^*(f_{\alpha_1})\dotsm
  a^*(f_{\alpha_l})\Omega$, where the notation is as in \eqref{ALPHA},
  and $a^*(f_j)$ and $\Omega$ are now creation \OP s and the Fock
  vacuum in $\FF_s(P_V(\ker V\+))$.
\end{Cor}
{\em Acknowledgement. } It is a pleasure to thank Dr.~M.~Schmidt for
discussions and for bringing refs.~\cite{Si,Se,M} to the author's
attention.

\end{document}